\documentclass[aps,prl,showpacs,twocolumn,superscriptaddress]{revtex4}
\usepackage{amsmath}
\usepackage{amssymb}
\usepackage{epsfig}
\usepackage{color}
\usepackage{subfigure}
\usepackage{graphicx,amsmath}

\begin{document}
\title{Scaling in Dynamic Susceptibility of Herbertsmithite and
Heavy-Fermion Metals}
\author{V. R. Shaginyan
\footnote{E-mail addresses:
vrshag@thd.pnpi.spb.ru}}
\affiliation{Petersburg Nuclear Physics Institute, Gatchina,
188300, Russia} \affiliation{Clark Atlanta University, Atlanta, GA
30314, USA}
\author{A. Z. Msezane}\affiliation{Clark Atlanta University,
Atlanta, GA 30314, USA}\author{K. G. Popov}\affiliation{Komi
Science Center, Ural Division, RAS, Syktyvkar, 167982,
Russia}\author{V. A. Khodel} \affiliation{Russian Research Center
Kurchatov Institute, Moscow, 123182, Russia} \affiliation{McDonnell
Center for the Space Sciences \& Department of Physics, Washington
University, St.~Louis, MO 63130, USA}

\begin{abstract}
We present a theory of the dynamic magnetic susceptibility of
quantum spin liquid. The obtained results are in good agreement
with experimental facts collected on herbertsmithite $\rm
ZnCu_3(OH)_6Cl_2$ and on heavy-fermion metals, and allow us to
predict a new scaling in magnetic fields in the dynamic
susceptibility. Under the application of strong magnetic fields
quantum spin liquid becomes completely polarized. We show that this
polarization can be viewed as a manifestation of gapped excitations
when investigating the spin-lattice relaxation rate.
\end{abstract}

\pacs{75.40.Gb, 64.70.Tg, 76.60.Es, 71.10.Hf\\ {\it Key Words}:
Quantum phase transitions; Strongly correlated spin liquid; Heavy
fermions; Dynamic magnetic susceptibility; Spin-lattice relaxation
rate}

\maketitle

\section{Introduction}

Landau Fermi liquid (LFL) theory is highly successful in the
condensed matter physics. The key point of this theory is the
existence of fermionic quasiparticles defining the thermodynamic,
relaxation and dynamic properties of the material. However,
strongly correlated Fermi systems  encompassing a variety of
systems that display behavior not easily understood within the
Fermi liquid theory and called non-Fermi liquid (NFL) behavior. A
paradigmatic example of the NFL behavior is represented by
heavy-fermion (HF) metals, where a quantum phase transition (QPT)
induces a transition between LFL and NFL \cite{loh,pr}. QPT can be
tuned by different parameters, such as the chemical composition,
the pressure, and the magnetic field. Magnetic materials, in
particular copper oxides and organic insulators, are interesting
subjects of study due to a quantum spin liquid (QSL) that can
emerge when they approach QPT and are cooled to low temperature
$T$. Exotic QSL is formed with such hypothetic particles as
fermionic spinons carrying spin $1/2$ and no charge. A search for
the materials is a challenge for condensed matter physics
\cite{bal}. The experimental studies of herbertsmithite $\rm
ZnCu_3(OH)_6Cl_2$ and the organic insulator $\rm
EtMe_3Sb[Pd(dmit)_2]_2$ have discovered gapless excitations,
analogous to excitations near the Fermi surface in HF metals,
indicating that $\rm ZnCu_3(OH)_6Cl_2$ and $\rm
EtMe_3Sb[Pd(dmit)_2]_2$ are the promising systems to investigate
their QPTs and QSLs
\cite{herb0,herb4,herb1,herb2,mil,herb3,herb,sl,sl1,sl2,prbr}. The
observed behavior of the thermodynamic properties of $\rm
ZnCu_3(OH)_6Cl_2$ strongly resembles that in HF metals since a
simple kagome lattice being strongly frustrated has a
dispersionless topologically protected branch of the spectrum with
zero excitation energy \cite{green,vol,prbr,eplh}. This indicates
that QSL formed by the ideal kagome lattice is located on the
ordered side of the fermion condensation quantum phase transition
(FCQPT) that is characterized by the presence of the spectrum with
zero excitation energy \cite{pr}. This observation allows us to
establish a close connection between QSL and HF metals whose HF
systems are located near FCQPT and, therefore, exhibiting an
universal scaling behavior \cite{pr,prbr,eplh}. As we are dealing
with the real 3D compound $\rm ZnCu_3(OH)_6Cl_2$ rather than with
the ideal 2D kagome lattice, we have to bear in mind that the
magnetic interactions and the presence of layers of nonmagnetic
$\rm Zn^{2+}$ ions separating magnetic kagome planes in the
substance can shift the QSL from the initial point, positioning it
in front of or behind FCQPT. Therefore, the actual location has to
be established by analyzing the experimental data. As a result, the
location coincides with that of HF metals, and turns out to be at
FCQPT \cite{prbr,eplh}, as it is shown in Fig. \ref{fig1}. Thus,
FCQPT can be considered as QPT of $\rm ZnCu_3(OH)_6Cl_2$ QSL and
both herbertsmithite and HF metals can be treated in the same
framework, so that QSL is composed of fermions and these with zero
charge and spin $\sigma=\pm1/2$ occupy the corresponding Fermi
sphere with the Fermi momentum $p_F$ \cite{pr,prbr,eplh}. The
ground state energy $E(n)$ is given by the Landau functional
depending on the quasiparticle distribution function $n_\sigma({\bf
p})$, where ${\bf p}$ is the momentum. In spite of numerous
experimental facts collected in measurements of inelastic neutron
scattering spectrum and spin-lattice relaxation rates on
herbertsmithite, a theoretical understanding of how the dynamical
spin susceptibility of QSL behaves on approaching QPT and how it is
affected by external parameters, such as the magnetic field, is
still missing.

In this letter we employ the Landau transport equation to construct
the dynamical spin susceptibility. We elucidate how the calculated
susceptibility is affected by magnetic field and describe
experimental facts collected on herbertsmithite and heavy-fermion
metals. The obtained results are in good agreement with the facts
and allow us to predict a new scaling emerging under the
application of magnetic field in the dynamic susceptibility. Taking
into account that QSL becomes completely polarized in strong
magnetic fields, we show that this polarization can be seen as the
presence of gapped excitations when investigating the spin-lattice
relaxation rate.

\section{Dynamic spin susceptibility of quantum spin liquid and heavy-fermion metals}

To construct the dynamic spin susceptibility $\chi({\bf
q},\omega,T)=\chi{'}({\bf q},\omega,T)+i\chi{''}({\bf q},\omega,T)$
as a function of momentum $q$, frequency $\omega$ and and
temperature $T$, we use the model of homogeneous HF liquid located
near FCQPT \cite{pr}. To deal with the dynamic properties of Fermi
systems, one can use the transport equation describing a slowly
varying disturbance $\delta n_{\sigma}({\bf q},\omega)$ of the
quasiparticle distribution function $n_0({\bf p})$, and $n=\delta
n+n_0$. We consider the case when the disturbance is induced by the
application of external magnetic field $B=B_0+\lambda B_1({\bf
q},\omega)$ with $B_0$ being a static field and $\lambda B_1$ a
$\omega$-dependent field with $\lambda\to0$. As long as the
transferred energy $\omega<qp_F/M^*<<\mu$, where $M^*$ is the
effective mass and $\mu$ is the chemical potential, the
quasiparticle distribution function $n({\bf q},\omega)$ satisfies
the transport equation \cite{PinNoz}
\begin{eqnarray}\label{TREQ}
&&({\bf qv_p}-\omega)\delta n_{\sigma}-{\bf qv_p}\frac{\partial
n_0}{\partial\varepsilon_p}\sum_{\sigma_1{\bf
p}_{1}}f_{\sigma,\sigma_1}({\bf pp}_{1})\delta n_{\sigma_1}({\bf
p}_{1})\nonumber\\&=&{\bf qv_p}\frac{\partial
n_0}{\partial\varepsilon_p}\sigma\mu_B(B_0+\lambda B_1).
\end{eqnarray}
Here $\mu_B$ is the Bohr magneton and $\varepsilon_p$ is the
single-particle spectrum. We assume that $B_0$ is finite but not so
strong to lead to the full polarization of the corresponding
quasiparticle band. In the field $B_0$, the two Fermi surfaces are
displaced by opposite amounts, $\pm B_0\mu_B$, and the magnetization
$\mathcal{M}=\mu_B(\delta n_+-\delta n_-)$, where the two spin
orientations with respect to the magnetic field are denoted by
$\pm$, and $\delta n_{\pm}=\sum_p \delta n_{\pm}({\bf p})$. The spin
susceptibility $\chi$ is given by $\chi=\partial\mathcal{M}/\partial
B_{|_{B=B_0}}$. In fact, the transport equation \eqref{TREQ} is
reduced to two equations which can be solved for each direction
$\pm$ and allows one to calculate $\delta n_{\pm}$ and the
magnetization. The response to the application of $\lambda B_1({\bf
q},\omega)$ can be found by expanding the solution of Eq.
\eqref{TREQ} in a power series with respect to $M^*\omega/qp_F$. As
a result, we obtain the imaginary part of the spin susceptibility
\begin{equation}\label{chi2}
\chi{''}({\bf q},\omega)=\mu_B^2\frac{\omega(M^*)^2}{2\pi
q}\frac{1}{(1+F^a_0)^2},
\end{equation}
where $F^a_0$ is the dimensionless spin antisymmetric quasiparticle
interaction \cite{PinNoz}. The interaction $F^a_0$ is found to
saturate at $F^a_0\simeq -0.8$ \cite{pfw,vollh1} so that
$(1+F^a_0)$ is positive. It is seen from Eq. \eqref{chi2} that the
second term is an odd function of $\omega$. Therefore, it does not
contribute to the real part $\chi'$ and forms the imaginary part
$\chi''$. Taking into account that at relatively high frequencies
$\omega\geq qp_F/M^*\ll\mu$ in the hydrodynamic approximation
$\chi'\propto 1/\omega^2$ \cite{forst}, we conclude that the
equation
\begin{equation}\label{chi3}
\chi({\bf
q},\omega)=\frac{\mu_B^2}{\pi^2(1+F^a_0)}\frac{M^*p_F}{1+i\pi
\frac{M^*\omega}{qp_F(1+F^a_0)}},
\end{equation}
produces the simple approximation for the susceptibility $\chi$ and
satisfies the Kramers-Kronig relation connecting the real and
imaginary parts of $\chi$.

To understand how can $\chi''$ and $\chi$ given by Eqs.
\eqref{chi2} and \eqref{chi3}, respectively, depend on temperature
$T$ and magnetic field $B$, we recall that near FCQPT point the
effective mass $M^*$ depends on $T$ and $B$, and is given by the
Landau equation (LE) \cite{pr,land}. The interaction function $F$
of LE is completely defined by the fact that the system has to be
at FCQPT. The sole role of $F$ is to bring the system to FCQPT,
where the Fermi surface alters its topology so that $M^*$ acquires
$T$ and $B$ dependencies \cite{pr,ckz,khodb}. At FCQPT, LE can be
solved analytically: At $B=0$, the effective mass depends on $T$
\begin{equation}
M^*(T)\simeq a_TT^{-2/3}.\label{MTT}
\end{equation}
At finite $T$, the application of magnetic field $B$ drives the
system to the LFL region with
\begin{equation}
M^*(B)\simeq a_BB^{-2/3}.\label{MBB}
\end{equation}
Here $a_T$ and $a_B$ are constants. At finite $B$ and $T$ near
FCQPT, the solutions of LE can be well approximated by a simple
universal interpolating function. The interpolation occurs between
the LFL ($M^*(T)\propto const)$ and NFL ($M^*(T)\propto T^{-2/3}$)
regions. It is convenient to introduce the normalized effective
mass $M^*_N$ and the normalized temperature $T_N$ dividing the
effective mass $M^*$ by its maximal values, $M^*_{\rm max}$, and
temperature $T$ by $T_{\rm max}$ at which the maximum occurs. The
normalized effective mass $M^*_N=M^*/M^*_{\rm max}$ as a function
of the normalized temperature $y=T_N=T/T_{\rm max}$ is given by the
interpolating function \cite{pr}
\begin{equation}M^*_N(y)\approx c_0\frac{1+c_1y^2}{1+c_2y^{8/3}}.
\label{UN2}
\end{equation}
Here $c_0=(1+c_2)/(1+c_1)$, $c_1$ and $c_2$ are fitting parameters,
making $M^*_N(y=1)=1$. Magnetic field $B$ enters LE only in the
combination $\mu_BB/k_BT$, making $k_BT_{\rm max}\simeq \mu_BB$
where $k_B$ is the Boltzmann constant \cite{ckz,pr}. Thus, in the
presence of magnetic fields the variable $y$ becomes $y=T/T_{\rm
max}\simeq k_BT/\mu_BB$. Since the variables $T$ and $B$ enter
symmetrically Eq. \eqref{UN2} is valid for $y=\mu_BB/k_BT$.
\begin{figure}[!ht]
\begin{center}
\includegraphics [width=0.47\textwidth]{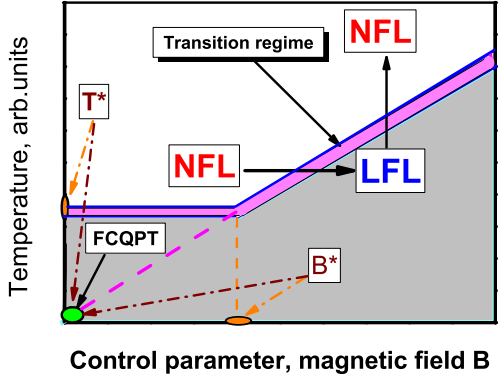}
\end{center}
\caption{(color online). $T-B$ phase diagram of QSL and HF liquid.
The vertical and horizontal arrows, crossing the transition region
depicted by the thick lines, show LFL-NFL and NFL-LFL transitions
at fixed $B$ and $T$, respectively. At temperatures $T<T^*$ and
magnetic field $B<B^*$ shown by the dash-dot arrows the effective
mass $M^*\simeq const$ and the system in the LFL region. The dash
line continuing the thick line represents the transition region
provided the system were located at FCQPT shown by the
arrow.}\label{fig1}
\end{figure}
Now we construct the schematic $T-B$ phase diagram of QSL and HF
liquid reported in Fig. \ref{fig1}. At $T=0$ and $B=0$ the system
can exactly be located at the FCQPT point without tuning with both
$T^*$ and $B^*$ are zero. It can also be shifted from the FCQPT
point by doping, pressure etc. In that case $T^*$ and $B^*$ become
finite so that at $T<T^*$ and $B<B^*$ the effective mass $M^*\simeq
const$. As seen from Fig. \ref{fig1}, at $T\simeq T^*$ and $B\simeq
B^*$ the transition region exhibits a kink, since $M^*$ is no
longer constant at rising $B$ and $T$. Magnetic field $B$ and
temperature $T$ play the role of the control parameters, driving it
from the NFL to LFL regions as shown by the vertical and horizontal
arrows. At fixed $B$ and increasing $T$ the system transits along
the vertical arrow from the LFL region to NFL one crossing the
transition region. On the contrary, at fixed $T$ increasing $B$
drives the system along the horizontal arrow from the NFL region to
LFL one.

\section{Scaling behavior of the dynamic susceptibility}

\begin{figure} [! ht]
\begin{center}
\includegraphics [width=0.47\textwidth]{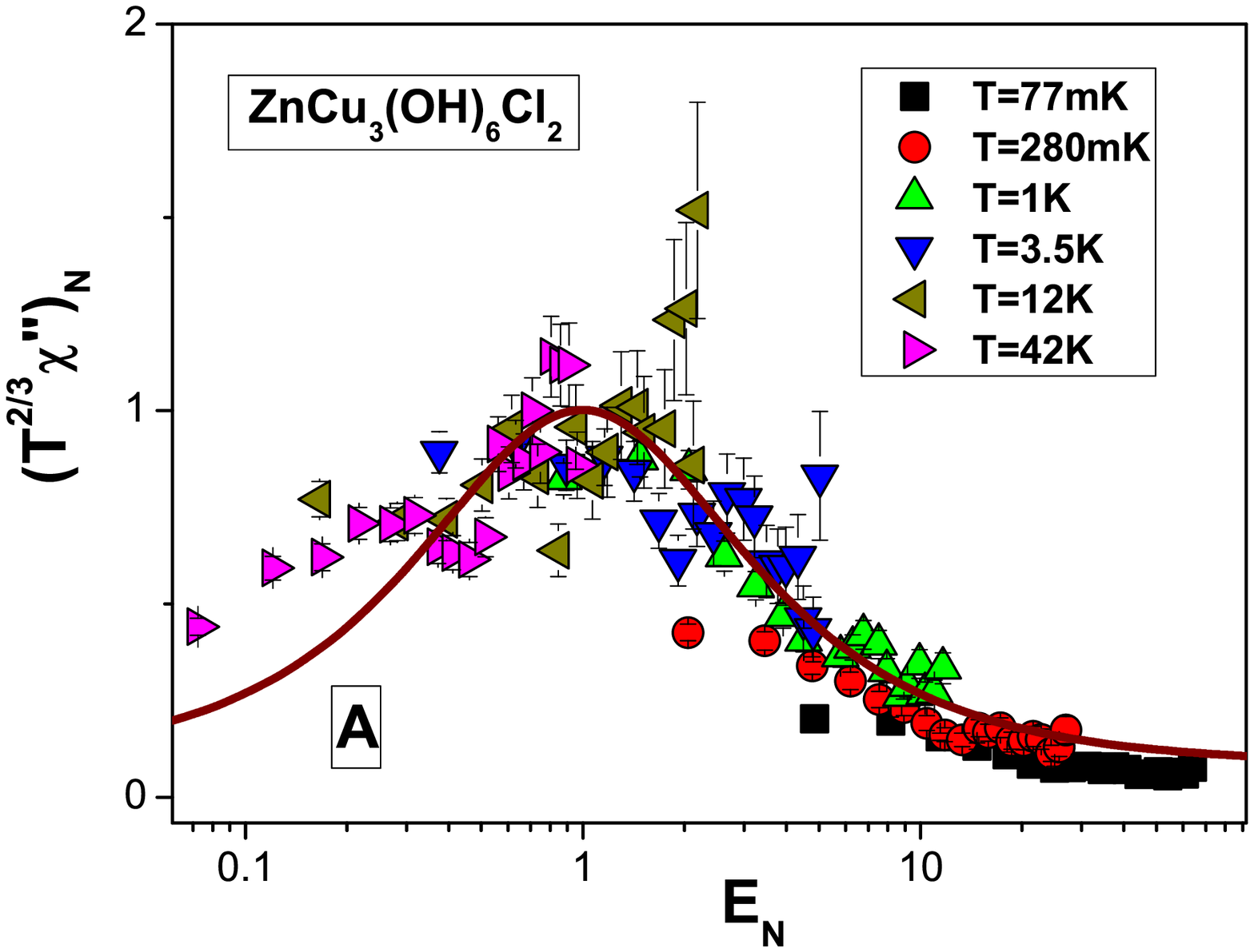}
\includegraphics [width=0.47\textwidth]{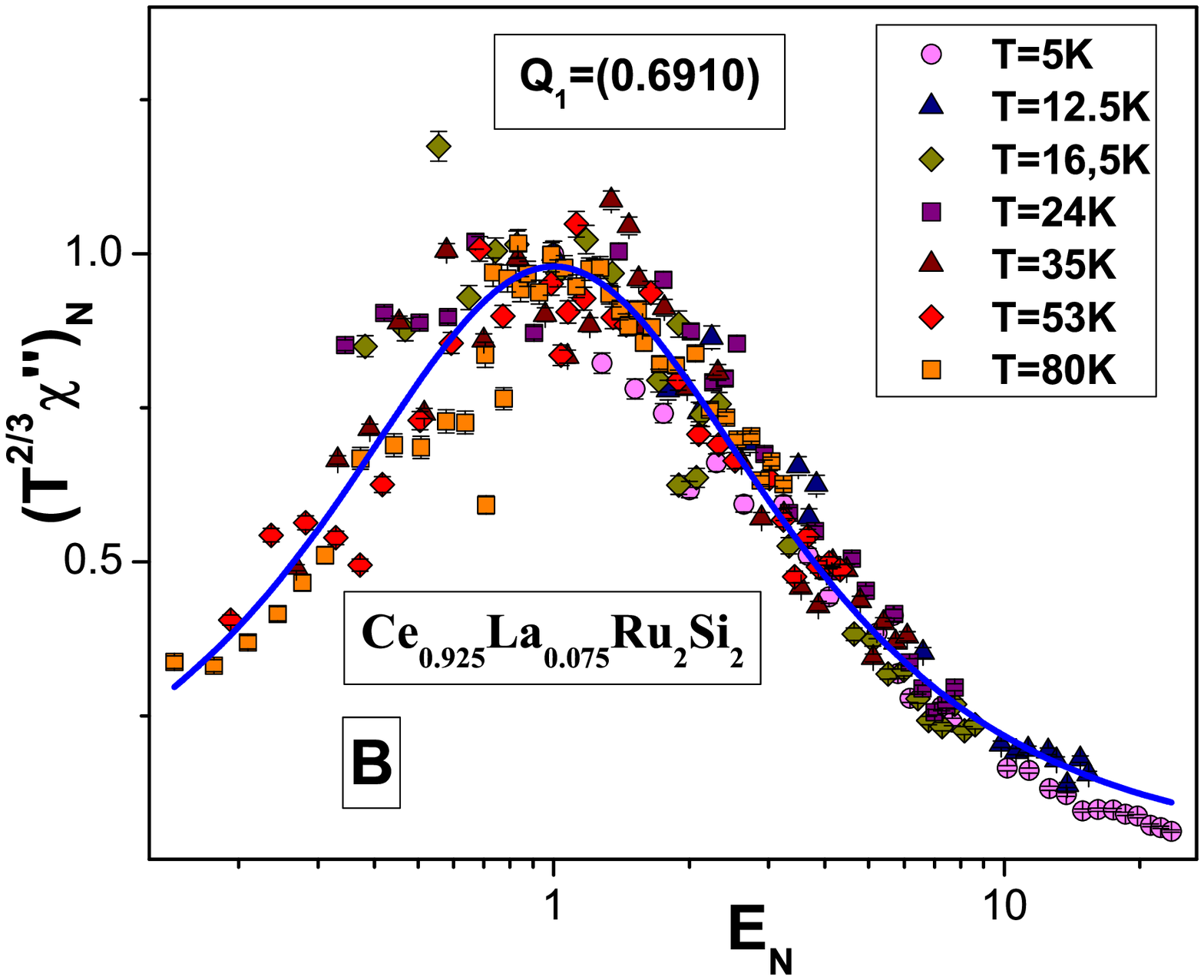}
\includegraphics [width=0.47\textwidth]{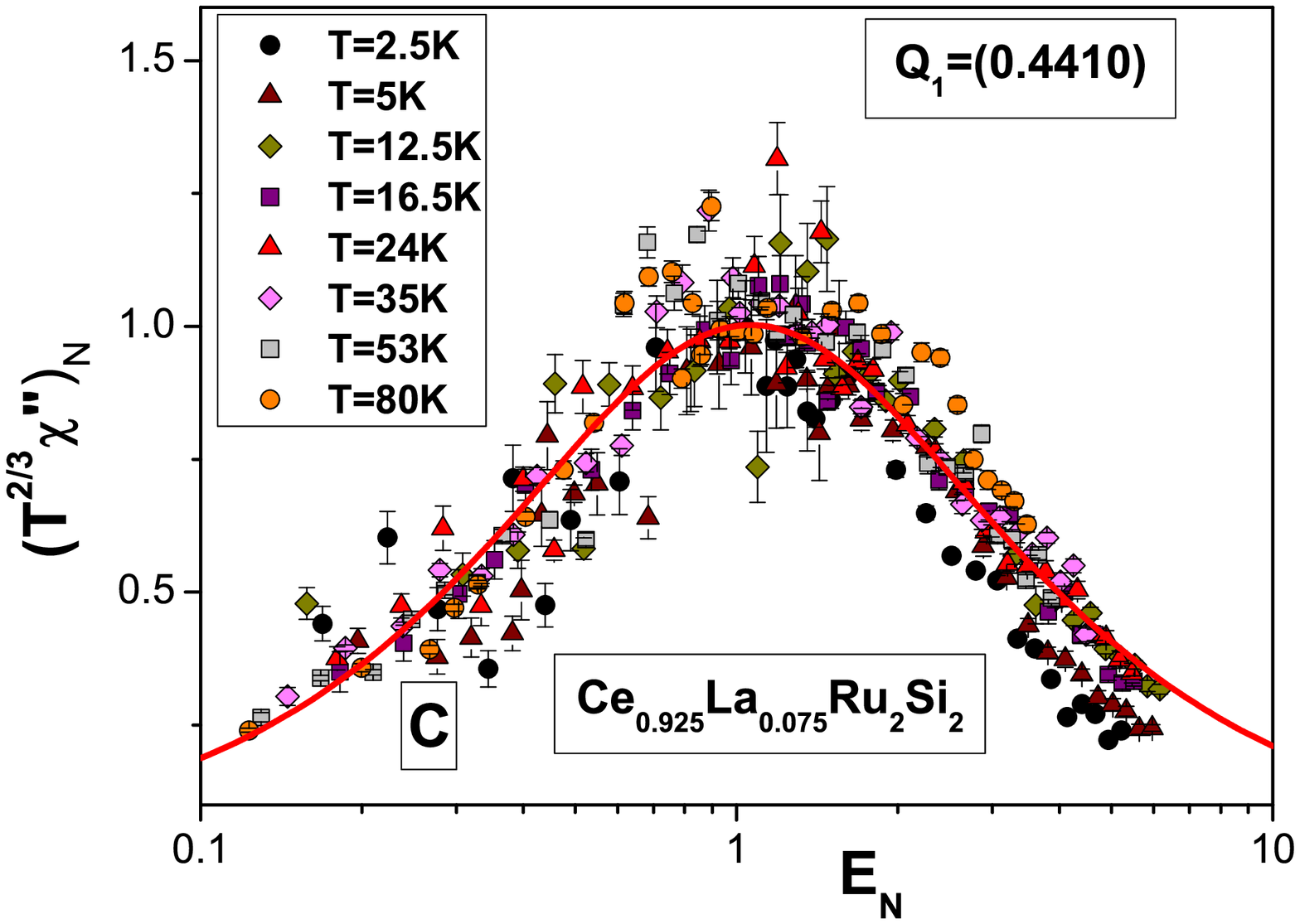}
\end{center}
\caption{(color online). The function $(T^{2/3}\chi'')_N$ plotted
against the unitless ratio $E_N=\omega/((k_BT)^{2/3}E_{\rm max})$.
The data extracted from measurements on $\rm ZnCu_3(OH)_6Cl_2$
obtained for $0.077<T<42$ K \cite{herb3}, Panel A, and on the HF
metal $\rm Ce_{0.925}La_{0.075}Ru_2Si_2$ obtained for $2.5<T <80$ K
at $Q_1$ \cite{prbh}, Panel B and C, collapse onto a single curve.
The solid curves are fits with the function given by Eq.
\eqref{SCHIN}.}\label{fig2}
\end{figure}

To elucidate a scaling behavior of $\chi$, we employ Eq.
\eqref{MTT} to describe the temperature dependence of $\chi$. It
follows from Eqs. \eqref{chi3} and \eqref{MTT} that
\begin{equation}\label{SCHI}
T^{2/3}\chi(T,\omega)\simeq \frac{a_1}{1+ia_2E}.
\end{equation}
Here $a_1$ and $a_2$ are constants absorbing irrelevant values and
$E=\omega/(k_BT)^{2/3}$. As a result, the imaginary part
$\chi''(T,\omega)$ satisfies the equation
\begin{equation}\label{SCHII}
T^{2/3}\chi''(T,\omega)\simeq\frac{a_3E}{1+a_4E^2},
\end{equation}
where $a_3$ and $a_4$ are constants. It is seen from Eq.
\eqref{SCHII} that $T^{2/3}\chi''(T,\omega)$ has a maximum
$(T^{2/3}\chi''(T,\omega))_{\rm max}$ at some $E_{\rm max}$ and
depends on the only variable $E$. Equation \eqref{SCHII} confirms
the scaling behavior of $\chi'' T^{0.66}$ experimentally
established in Ref. \cite{herb3}. As it was done for the effective
mass when constructing \eqref{UN2}, we introduce the dimensionless
function $(T^{2/3}\chi'')_{N}=T^{2/3}\chi''/(T^{2/3}\chi'')_{\rm
max}$ and the dimensionless variable $E_N=E/E_{\rm max}$, and Eq.
\eqref{SCHII} is modified to read
\begin{equation}\label{SCHIN}
(T^{2/3}\chi'')_N\simeq\frac{b_1E_N}{1+b_2E_N^2},
\end{equation}
with $b_1$ and $b_2$ are fitting parameters which are to adjust the
function on the right-hand side of Eq. \eqref{SCHIN} to reach its
maximum value 1 at $E_n=1$. We predict that if measurements of
$\chi''$ are taken at fixed T as a function of $B$, then taking
into account Eq. \eqref{MBB}, we again obtain that the function
$B^{2/3}\chi''(E)$ exhibits the scaling behavior with
$E=\omega/(\mu_BB^{2/3})$. If the system is placed at FCQPT, the
scaling described above is valid down to lowest temperatures. When
the system is shifted from FCQPT then $T^*$ and $B^*$ are finite
and the scaling is violated in the LFL region and recovered in the
NFL one at $T>T^*$ or $B>B^*$ as seen from Fig. \ref{fig1}.

In Fig. \ref{fig2} consistent with Eq. \eqref{SCHIN}, the scaling
of the normalized dynamic susceptibility $(T^{2/3}\chi'')_N$
extracted from the inelastic neutron scattering spectrum of both
herbertsmithite \cite{herb3}, Panel A, and $\rm
Ce_{0.925}La_{0.075}Ru_2Si_2$, Panel B and C, \cite{prbh} is
displayed. The scaled data collapse fairly well onto a single curve
over almost three decades of $E_N$. It is seen that our
calculations shown by the solid curves are overall in good
agreement with the experimental facts. We note that, as seen from
Fig. \ref{fig2}, Panel C, the data taken at 2.5 K are at variance
with the scaling behavior shown by the solid curve. We suggest that
$\rm Ce_{0.925}La_{0.075}Ru_2Si_2$ is slightly shifted from FCQPT
as shown in Fig. \ref{fig1} and at sufficiently low temperatures
$T<T^*$ the scaling is violated \cite{prbh,stegr}, while $\rm
ZnCu_3(OH)_6Cl_2$ is near that point. Some remarks on a role of
both the disorder and the anisotropy are in order. The anisotropy
is supposed to be related to the Dzyaloshinskii-Moriya interaction,
exchange anisotropy, or out-of-plane impurities. Measurements of
the susceptibility on the single crystal of herbertsmithite have
shown that it closely follows that measured on a powder sample
\cite{herb}. At low temperatures $T\lesssim70$ K, the
single-crystal data do not show magnetic anisotropy \cite{herb}.
These confirm that the stoichiometry, disorder and anisotropy do
not contribute significantly to the results at relatively low
temperatures. Moreover, the scaling behavior of the thermodynamic
functions of herbertsmithite is the intrinsic feature and has
nothing to do with the impurities \cite{eplh}. These observations
are in agreement with a general consideration of scaling behavior
of HF metals \cite{pr}.

\section{Spin-lattice relaxation rate of quantum spin liquid}

\begin{figure} [! ht]
\begin{center}
\includegraphics [width=0.47\textwidth]{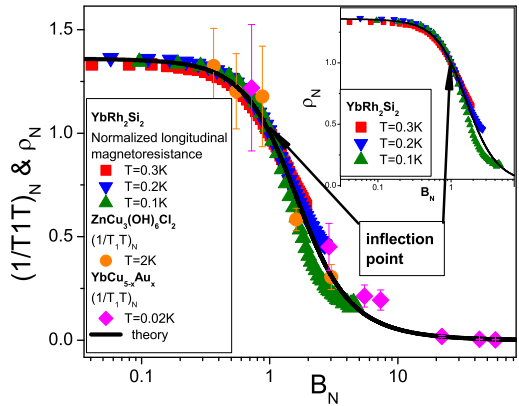}
\end{center}
\caption{(color online). Magnetic field dependence of normalized
(see text for details) muon spin-lattice relaxation rate
$(1/T_1T)_N$ extracted from measurements on ${\rm
{YbCu_{4.4}Au_{0.6}}}$ \cite{carr} and $\rm ZnCu_3(OH)_6Cl_2$
\cite{imai} along with the normalized longitudinal
magnetoresistance $\rho_N$  versus normalized magnetic field $B_N$.
Our calculations are shown by the solid line. The arrows indicate
the inflection points. Inset: $\rho_N$ versus $B_N$, $\rho_N$ is
extracted from measurements on $\rm YbRh_2Si_2$ at different
temperatures \cite{steg} listed in the legend. The solid curve
represents our calculations. }\label{fig3}
\end{figure}

Consider the effect of $B$ on the spin-lattice relaxation rate
$1/T_1T$ determined by $\chi''$ given by Eq. \eqref{chi2}
\begin{equation}\label{chi11}
\frac{1}{T_1T}=\frac{3}{4\mu_B^2}\sum _{\bf q}A_{\bf q}A_{-{\bf
q}}\frac{\chi''({\bf q},\omega)}{\omega}{|_{\omega\to0}}\propto
(M^*)^2,
\end{equation}
where $A_{\bf q}$ is the hyperfine coupling constant of the muon
(or nuclei) with the spin excitations at wave vector $\bf q$
\cite{korr,kn,pr}. Figure \ref{fig3} and the inset display the
normalized $(1/T_1T)_N$ and the normalized longitudinal
magnetoresistance $\rho_N$ at fixed temperature versus the
normalized magnetic field $B_N$. It is seen from  Fig. \ref{fig3}
that the magnetic field progressively reduces $1/T_1T$ and the
longitudinal magnetoresistance (LMR), and these as a function of
$B$ possess an inflection point at $B=B_{inf}$ shown by the arrow.
The normalized LMR obeys the equation \cite{pr}
\begin{equation}\label{rn}
\rho_N(B_N)=\frac{\rho(B_N)-\rho_0}{\rho_{inf}}=\left(\frac{1}{T_1T}\right)_N
=(M_N^*)^2,
\end{equation}
where $\rho_0$ is the residual resistance, $\rho_{inf}$ is LMR
taken at the inflection point, $\rho$ is LMR,  and $B_N=B/B_{inf}$.
We normalize $(1/T_1T)$ and LMR by their values at the inflection
point, and the magnetic field is normalized by $B_{inf}$. In
accordance with the phase diagram \ref{fig1}, at $B>B_{inf}$, as
seen from Fig. \ref{fig3}, QSL enters the LFL region with
$B$-dependence of the effective mass defined by Eq. \eqref{MBB}. It
follows from Eqs. \eqref{chi11} and \eqref{rn} that
$(1/T_1T)_N=\rho_N=(M^*_N)^2$ where $(M^*_N)^2$ is defined by Eq.
\eqref{UN2} which shows that different strongly correlated Fermi
systems are to exhibit the same scaling of $(M^*_N)^2$. It is seen
from Fig. \ref{fig3} and from the inset, that $\rm
YbCu_{5-x}Au_{x}$, herbertsmithite $\rm ZnCu_3(OH)_6Cl_2$ and $\rm
YbRh_2Si_2$ demonstrate the similar behavior of $(M^*_N)^2$
resulting in the scaling of LMR and $1/T_1T$. Thus, Eqs.
\eqref{chi2}, \eqref{chi11} and \eqref{rn} determine the close
relationship existing between the quite different dynamic
properties and different strongly correlated Fermi systems such as
QSL and HF metals, revealing their scaling behavior at FCQPT.

We note that one may be confused when applying Eq. \eqref{chi11} to
describe $(1/T_1T)$ in strong magnetic fields. In that case both
QSL and HF metals become fully polarized due to Zeeman splitting
\cite{korr,kn,prbr,epl}. As a result, one subband becomes empty,
while the energy $\varepsilon_F$ of spinons at the Fermi surface of
the other subband lies below the chemical potential $\mu$ formed by
the magnetic field $B_0$. It follows from Eq. \eqref{TREQ} that
$\chi''=0$ and Eq. \eqref{chi11} is not valid. The difference
$\delta=\mu-\varepsilon_F$ can be viewed as a gap that makes
$1/T_1T\propto \exp{-(\delta/k_BT)}$. At temperatures
$k_BT\sim\delta$, the subbands are populated by spinons and the
validity of Eq. \eqref{chi11} is restored. Thus, $\delta$ can be
interpreted as the presence of gapped excitations. On the other
hand, if there were the gapped excitations, then the heat capacity
demonstrates the exponential decay rather than a linear
$T$-dependence at low temperatures. Analysis based on experimental
data shows the presence of linear $T$-dependence even under the
application of high magnetic fields \cite{prbr}, while recent
measurements on $\rm ZnCu_3(OH)_6Cl_2$ of $1/T_1T$ suggest the
gapped excitations \cite{qslg}. To clarify whether the gapped
excitations would occur in $\rm ZnCu_3(OH)_6Cl_2$, an accurate
experimental measurement in magnetic fields of the low temperature
heat capacity is necessary.

\section{Conclusions}

We have presented a theory of the dynamic magnetic susceptibility
of quantum spin liquid, and elucidated how the calculated
susceptibility is affected by magnetic field and describe
experimental facts collected on herbertsmithite and heavy-fermion
metals. The obtained results are in good agreement with
experimental facts collected on both herbertsmithite $\rm
ZnCu_3(OH)_6Cl_2$ and on heavy-fermion metals, and allow us to
conclude that the dynamic magnetic susceptibility of
herbertsmithite is similar to that of heavy-fermion metals. Thus,
herbertsmithite can be viewed as a new type of strongly correlated
electrical insulator that possesses properties of heavy-fermion
metals with one exception: it resists the flow of electric charge.
We have also predicted a new scaling in magnetic fields in the
dynamic susceptibility emerging under the application of magnetic
field. Taking into account that under the application of strong
magnetic fields quantum spin liquid becomes completely polarized,
we have shown that this polarization can be viewed as a
manifestation of gapped excitations when investigating the
spin-lattice relaxation rate.

\section{Acknowledgements}

This work was supported by U.S. DOE, Division of Chemical Sciences,
Office of Basic Energy Sciences, Office of Energy Research, and
AFOSR.


\begin{thebibliography}{99}

\bibitem{loh} H.v. L\"ohneysen, A. Rosch, M. Vojta, P. W\"olfle,
\rmp 79 (2007) 1015.

\bibitem{pr} V. R. Shaginyan, M. Ya. Amusia, A. Z. Msezane,
K. G. Popov,  Phys. Rep. 492 (2010) 31.

\bibitem{bal} L. Balents, Nature 464 (2010) 199.

\bibitem{herb0} M. P. Shores, E. A. Nytko, B. M. Bartlett, D. G.
Nocera, J. Am. Chem. Soc. 127 (2005) 13462.

\bibitem{herb1} J. S. Helton, et al., \prl 98 (2007) 107204.

\bibitem{herb2} M. A. deVries, K. V. Kamenev, W. A. Kockelmann,
J. Sanchez-Benitez, A. Harrison \prl 100 (2008) 157205.

\bibitem{herb3} J. S. Helton, et al., \prl 104 (2010) 147201.

\bibitem{herb} T. H. Han, et al., \prb 83 (2011) 100402(R).

\bibitem{herb4} F. Bert, P. Mendels, J. Phys. Soc. Jpn. 79 (2010) 011001.

\bibitem{mil} F. Mila, \prl 81 (1998) 2356.

\bibitem{sl} S. S. Lee, P. A. Lee, \prl 95 (2005) 036403.

\bibitem{sl1} M. Yamashita, et al., Science 328 (2010) 1246.

\bibitem{sl2} Y. Ran, M. Hermele, P. A. Lee, X. G. Wen, \prl 98 (2007)
117205.

\bibitem{prbr} V. R. Shaginyan, A. Z. Msezane, K. G. Popov,
\prb 84 (2011) 060401(R).

\bibitem{green} D. Green, L. Santos, C. Chamon, \prb 82 (2010) 075104.

\bibitem{vol} T. T. Heikkila, N. B. Kopnin, G. E. Volovik,
JETP Lett. 94 (2011) 233.

\bibitem{eplh} V. R. Shaginyan, A. Z. Msezane, K. G. Popov, G. S. Japaridze,
V. A. Stephanovich, Europhys. Lett. 97 (2012) 56001.

\bibitem{PinNoz} D. Pines, P. Nozi\'eres,
Theory of Quantum Liquids, Benjamin, New York, 1966.

\bibitem{pfw} M. Pfitzner, P. W\"olfle,  Phys. Rev. B 33 (1986) 2003.

\bibitem{vollh1} D. Vollhardt, P. W\"olfle, P. W. Anderson, Phys.
Rev. B 35 (1987) 6703.

\bibitem{forst} D. Forster, Hydrodynamic Fluctuations, Broken
Symmetry, and Correlation Functions, W. A. Benjamin, Inc. 1975.

\bibitem{land} L. D. Landau, Sov. Phys. JETP 3 (1956) 920.

\bibitem{ckz} J. W. Clark, V. A. Khodel, M. V. Zverev,
\prb 71 (2005) 012401.

\bibitem{khodb} V. A. Khodel, J. W. Clark, M. V. Zverev,
\prb 78 (2008) 075120.

\bibitem{prbh} W. Knafo, et al., \prb 70 (2004) 174401.

\bibitem{stegr} O. Stockert, F. Steglich,
Annu. Rev. Condens. Matter Phys. 2 (2011) 79.

\bibitem{korr} J. Koringa, Physica 16 (1950) 601.

\bibitem{kn} T. Moriya, Spin Fluctuations in Itinerant Electron
Magnetism, Springer, Berlin, 1985.

\bibitem{carr} P. Carretta, R. Pasero, M. Giovannini, C. Baines,
\prb 79 (2009) 020401(R).

\bibitem{imai} T. Imai, E. A. Nytko, B.M. Bartlett,
M. P. Shores, D. G. Nocera, \prl 100 (2008) 077203.

\bibitem{steg}  P. Gegenwart, et al., Science 315 (2007) 969.

\bibitem{epl} V. R. Shaginyan, et al., Europhys. Lett. 93 (2011) 17008.

\bibitem{qslg} M. Jeong, et al., \prl 107 (2011) 237201.

\end{thebibliography}
\end{document}